# The Influence of Space Environment on the Evolution of Mercury

Stefano Orsini[1], Valeria Mangano[1], Alessandro Mura[1], Diego Turrini[1],

Stefano Massetti[1], Anna Milillo[1], Christina Plainaki[1]

(1) INAF-IAPS, Istituto di Astrofisica e Planetologia Spaziali, Roma, Italy

**ABSTRACT**:

Mercury, due to its close location to the Sun, is surrounded by an environment whose conditions may be considered as 'extreme' in the entire Solar System. Both solar wind and radiation are stronger with respect to other Solar System bodies, so that their interactions with the planet cause high emission of material from its surface. Moreover, the meteoritic precipitation plays a significant role in surface emission processes. This emitted material is partially lost in space. Although under the present conditions the surface particles loss rate does not seem to be able to produce significant erosion of the planetary mass and volume, the long-term effects over billions of years should be carefully considered to properly understand the evolution of the planet. In the early stages, under even more extreme conditions, some of these processes were much more effective in removing material from the planet's surface. This study attempts to provide a rough estimation of the material loss rate as a function of time, in order to evaluate whether and how this environmental effect can be applied to understand the Hermean surface evolution. We show that the most potentially effective Sun-induced erosion process in early times is a combination of ion sputtering, photon stimulated desorption and enhanced diffusion, which could have caused the loss of a surface layer down to a depth of 20 m, as well as a relevant Na depletion.

## 1. Introduction

New space missions devoted to the study of planet Mercury have been planned. Actually, the NASA MESSENGER mission (Solomon et al., 2008) has already operated along its orbital path around the planet, and it is presently still collecting data. Another mission, the ESA-JAXA BepiColombo (Benkhoff et al., 2010), will be launched in 2016-2017. Because of these missions, the scientific community has been solicited to produce studies to better understand the Mercurian features and to maximise the return of the present and future in situ measurements. Mercury, because of its close distance to the Sun, is exposed to the highest solar wind density and UV flux in the Solar System. This causes high release of material from its surface, due to processes like photon-stimulated desorption (PSD) linked to enhanced diffusion (ED) (e.g: Wurz and Lammer, 2003; Killen et al. 2007).), thermal desorption (TD) and ion sputtering (IS). Moreover, the meteoritic precipitation causes meteoritic impact vaporization (MIV), which seems to play a significant role among emission processes as well (Morgan 1988, Cremonese et al., 2005, 2006; Borin et al, 2009, 2010).

Depending on the typical energy involved in these processes and transferred to the released particles, the emitted material may fall back to the surface or be lost in space. Under the present conditions such a loss rate seems not to be able to produce a significant erosion of the planetary surface; nevertheless, the long-term effects over billions of years should be carefully considered to understand if they could affect the surface evolution of the planet in the context of the evolution of the Solar System. Following the timeline of the Solar System's history drawn by Coradini et al. (2011), Mercury formed in the Primordial Solar System: a 0.5-1 Ga long phase extending from the dispersal of the nebular gas of the Solar Nebula (i.e. about 10 Ma after the condensation of the Ca-Al-rich inclusions (present in chondritic meteorites) to the end of the Late Heavy Bombardment (LHB, i.e., between 3.9 and 3.7 Ga ago). The LHB marked the transition to the Modern Solar System, a phase lasting until now and characterized by a less violent evolution of the Solar System and by more regular, secular processes (Coradini et al., 2011). Before proceeding it is important to point that, while in the following we will use the LHB as a landmark over the life of the Solar System, the actual duration and intensity of this event are still a matter of debate: we refer interested readers to Fassett & Minton (2013), Geiss & Rossi (2013) and references therein for a more complete discussion on the subject. The transition between the Primordial and the Modern Solar System has been proposed to also imply a change in the population of impactors, and thus in the cratering process of the planetary surfaces in the inner Solar System, both in terms of impact rate and size distribution of the impactors (Strom et al. 2005). 'Population 1' impactors, responsible for the LHB, were ejected from the asteroid belt by size-independent dynamical processes (plausibly, the sweeping of orbital resonances with the giant planets, Gomes et al., 2005) while 'Population 2' impactors, responsible for the cratering across the Modern Solar System phase, are injected in the inner Solar System by size-dependent processes (plausibly, the combined action of Yarkovsky effect and resonance crossing, Farinella & Vokrouhlicky, 1999; Morbidelli & Vokrouhlicky, 2003). According to Strom et al. (2005), the impacts caused by 'Population 1' could have erased the cratering record on the planetary surfaces in the inner Solar System: this would imply that we cannot use the cratering record of the terrestrial planets to investigate time periods earlier than about 4 Ga ago. However, before and across the LHB the Sun was still characterized by a much more intense activity, both in terms of solar wind and radiation, and also the meteoritic bombardment was likely more intense due to the higher production rates of dust and collisional shards in the asteroid belt because of its greater population (e.g.: Bottke et al., 2005a,b and Minton & Malhotra, 2010). Under these more extreme conditions, the previously mentioned erosive processes were much more effective in removing material from the planet's surface.

This study attempts to provide a first rough estimation of the loss rate induced by the Sun and meteoritic impacts as a function of time, in order to evaluate whether and how these environmental features were most effective and their importance for understanding the surface evolution of Mercury. The crustal surface erosion induced by the known loss processes is estimated for the earlier phases of the evolution of the Solar System through extrapolation from the present conditions.

The Mercurian surface evolution induced by the Sun is determined by applying the model by Mura et al. (2007), capable of deriving the exospheric profiles (of both gravitationally bound and escaping particles) as a function of external input parameters (surface composition, solar and space conditions, etc.).

This paper is organized as follows: in Section 2 we speculate on the characteristics of the solar radiation and solar wind throughout their historical evolution. In Section 3, we describe the processes able to deplete the surface of Mercury. In Section 4, we study the effectiveness of these release processes for the surface erosion: (i) a refined IS computation is performed, given the solar wind density and velocity profiles versus time, since early phases; (ii) the erosion induced by meteoritic impacts is also estimated, by considering both dust and bigger bodies ejected from the Main Belt; (iii) finally, we estimate the loss rates possibly induced by volatiles (mainly Na, but possibly other elements like e.g. K) depletion due to the much stronger solar radiation occurring in early phases. In Section 5 the implications of the described processes in terms of surface evolution are discussed. In Section 6, the major results are recapped.

## 2. Evolution of solar radiation and solar wind

In this section we discuss the temporal evolution of the environment around Mercury, and in particular the variation of those parameters that may affect and erode the surface: UV flux (Figure 1), solar wind velocity and density (Figures 2 and 3). The current understanding is based on studies that analyse the present status of a large sample of Sun-like stars, whose conditions correspond to different evolution phases, so that the historical profile of the solar emission can be derived. Both temperature and global luminosity increased vs. time (Guinan and Ribas, 2002), hence, like it happens today, in early phases the particle emission due to thermal desorption was not capable of producing significant erosion rates.

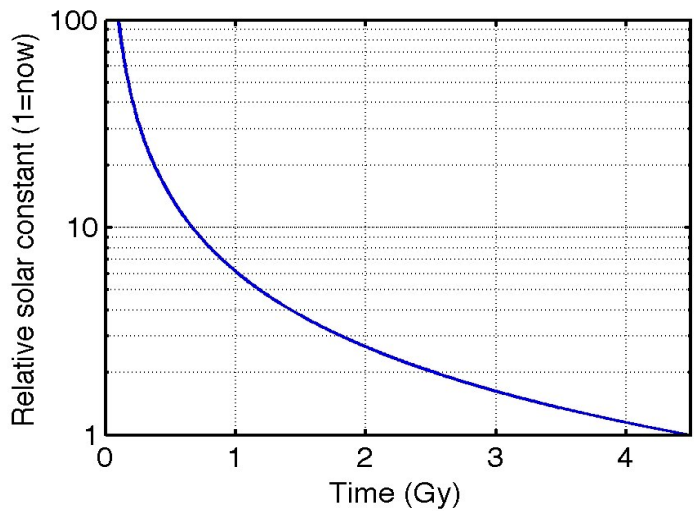

*Figure 1.* Solar UV flux vs. age of the Sun, normalized to the present situation. Data taken from Ribas et al. (2005), averaged over 100-360 Å wavelengths.

The UV radiation at 100-360 Å wavelengths is able to desorb Na and K from the planet's surface through PSD (Killen et al., 2007). As shown in Figure 1, this radiation evolves versus time, being much more intense in early phases. Over shorter time scales, also fluctuations related to solar cycle should be considered, but they are ineffective over the whole age of the Solar System. Since the typical energy range of the PSD process allows part of the ejecta to be lost in space, it follows that PSD could have been responsible for mass loss by Mercury during early phases: however, it is unclear whether this mass loss could result in a surface erosion as in the case of the other processes under study (see Sect. 3.2 for a detailed discussion). It is worthwhile to notice that PSD is a very selective process, so that only volatiles not strongly bounded into the surface minerals can be extracted and possibly lost in space. At present times, such a process may only marginally influence the whole particle escape budget. Actually, Na and K are the only exospheric volatile components of planetary origin observed to be relevant at Mercury (Potter et al, 2002: Leblanc and Johnson, 2003), since the observed He and H are probably a mixture of solar and planetary origin (e.g.: Milillo et al., 2010).

Moreover, both solar wind velocity (Figure 2) and density (Figure 3) decreased versus time (see Lundin et al., 2007, and references therein), so that, with respect to present time, in early phases the first was a factor 4 higher, and the second was a factor $10^2 - 10^3$ higher. Ion precipitation on the planet's surface generates material emission through IS processes, which produces a significant portion of material able to escape from Mercury (Wurz and Lammer, 2003; Mura et al., 2006). This effect, depending on solar wind density and velocity, should be considered as a potential source of surface erosion (Albarède, 2009). In fact, in normal conditions ions may strike the surface in the dayside cusps, but during extreme tail loading events, as seen by MESSENGER (Slavin et al., 2010), the entire dayside may be exposed to the solar wind, sometimes extending to the nightside too.

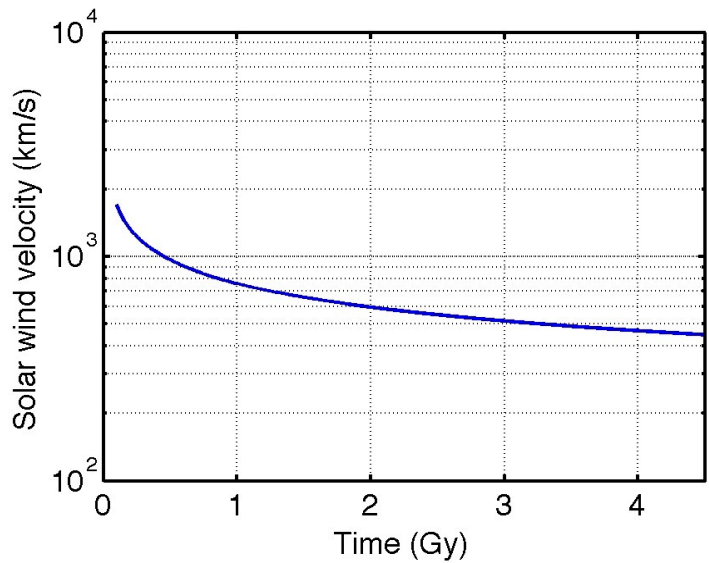

*Figure 2.* *Evolution of the solar wind velocity as a function of solar age. Data obtained from Kulikov et al.: 2006.*

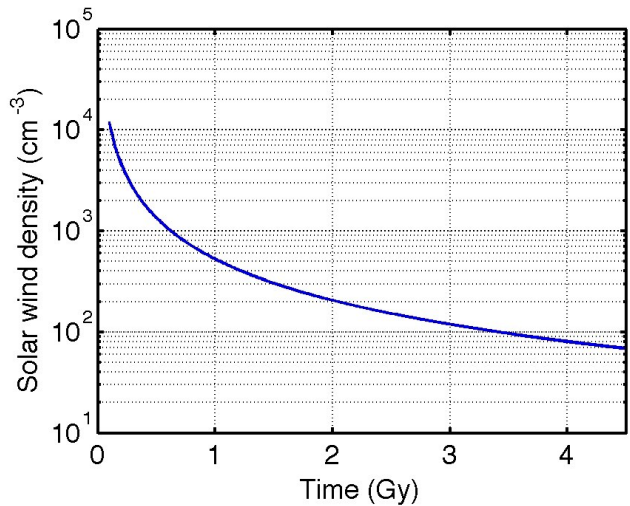

*Figure 3.* *Solar wind density as a function of the Sun's age, plotted according to Kulikov et al., 2006 (average solar wind density case, rescaled to Mercury's distance from the Sun).*

### 3. Release processes at the surface of Mercury

As already stated in Section 1, the surface of Mercury is eroded through several processes, like IS, PSD, and MIV: photons, ions, electrons or micrometeorites precipitating onto the surface transmit energy in the impact and produce the release of bound particles. Different studies (i.e. Wurz and Lammer, 2003; Mura et al., 2007) have shown that the particles extracted by TD do not have sufficient energy to escape from the planet; hence, this process may be assumed to be negligible in terms of surface erosion.

*3.1 Ion sputtering*

The IS process is defined as the removal of a part of atoms or molecules from a solid surface due to the interaction of a projectile ion with target electrons and nuclei, as well as secondary cascades of collisions between target atoms (Sigmund, 1969). If the impact energy $E_i$ is high enough, a surface particle may be extracted, thus resulting in the erosion of the surface. Although IS is in general a stoichiometric process, at Mercury the extraction coefficient (yield) does not vary significantly among different species and, hence, the surface is eroded almost uniformly. The distribution function ($f_S$) of the ejection energy ($E_e$) usually peaks at few eVs (Siegmund, 1969; Sieveka and Johnson, 1984). The resulting neutral differential flux for each released species is given by:

$$\frac{d\Phi_n}{dE_e} = Y_i c \int_{E\min}^{E\max} \frac{d\Phi_I}{dE_i} f_S(E_e, E_i) dE_i \qquad (1)$$

where $Y_i$ is the sputtering yield corresponding to solar wind ions of type i (i=1 corresponds to impacting $H^+$ and i=2 to heavier ions), $c$ is the surface relative abundance of the atomic species considered, and $\Phi_i$ is the ion flux of solar wind ions of type *i*. The yield $Y$ is not calculated for every single energy, as in this study this accuracy is not necessary; $Y$ is 10% on average, as in Mura et al., 2007, which is a conservative value with respect to previously used values (15% in Wurz et al., 2003). The yield reduction due to regolith porosity (Cassidy and Johnson, 2005) is not considered here, but it is compensated by the heavy ion component of the solar wind, that may contribute to this process as well (Baragiola et al., 2003); in fact, even if the flux is lower, the yield is larger.

Solar wind protons are expected to precipitate in the dayside cusps of Mercury (e.g.: Massetti et al., 2003, Kallio and Janhunen, 2003). The intensity and shape of the $H^+$ flux depends on the magnetospheric configuration, the interplanetary magnetic field (IMF) and the solar wind velocity and density (Kabin et al., 2000; Sarantos et al. 2001, Kallio and Janhunen, 2003, Massetti et al., 2003). More recently, the MAG and FIPS measurements from MESSENGER confirmed the presence of $H^+$ and $Na^+$ fluxes directed toward the surface in the Northern hemisphere (Winslow et al., 2012; Raines et al. 2013). Winslow et al. (2012) estimated an $H^+$ bombardment rate of $10^{24}$ $s^{-1}$ over an area of $5\times10^{11}$ $m^2$ below the Northern cusp. Furthermore, the recent MESSENGER observations indicate a northward offset of about 400 km of the internal magnetic field (Anderson et al. 2011). Hence, the weaker field in the southern cusp is expected to result in a 4 times greater precipitation rate (Winslow et al., 2012), and the H$^+$ flux onto the surface of Mercury may exceed values of $10^9$ cm$^{-2}$ s$^{-1}$ (Mura et al., 2005).

In a model spanning over several billion years, we assume that it is not worth to reconstruct accurately all reliable configurations. To reconstruct the H$^+$ flux onto the planetary surface we apply a Monte-Carlo model of proton circulation in Mercury's dayside, as in Mura et al. (2009), assuming that only solar wind velocity and density change over time (see Figures 2 and 3). Similarly, in our simulation the IMF *x*, *y* and *z* components are assumed to be constant (-15, 10, -10 nT), as we are not specifically interested in the possible evolution of the location where plasma precipitates, rather than in the global effect of such a process on the erosion rate. According to this approach, any short time scale variation is not effective in the model computations.

*3.2 Photon-stimulated desorption and volatiles abundance*

The surface of Mercury is exposed to an intense flux of photons; those of sufficiently high energy (UV or shorter wavelengths) may extract neutral atoms from the planetary surface. This process is very effective for the extraction of volatile species, like Na and K in the case of Mercury, while it is not efficient for refractory species (e.g.: Killen et al. 2007). For Na, the net flux from a surface containing this element can be as high as $10^9$ cm$^{-2}$ s$^{-1}$, and it is given by the following formula:

$$\Phi_n = Nc \underbrace{\int \Phi_\gamma(E)\sigma(E)dE}_{A} = NcA \qquad (2)$$

where $\Phi_n$ is the neutral flux, $\Phi_\gamma(E)$ is the differential photon flux function of energy *E*, $\sigma(E)$ is the cross-section as a function of E, *N* is the surface numerical density and *c* is the relative composition of the considered neutral species (for Na, respectively: $7.5 \cdot 10^{14}$ cm$^{-2}$ Killen et al., 2001 and 2.5%, Evans et al., 2012). Yakshinskiy and Madey (1999) report a study on the lunar surface: they experimentally found that, below approximately 250 nm (hν > 5 eV), photons can remove Na from a SiO$_2$ surface at 250 K with a cross-section of 1 to 3 $10^{-20}$ cm$^{-2}$, provided that free Na is available on the uppermost layer. In our model the factual rate is calculated by taking into account either the PSD efficiency or the ED rate as in Killen et al. (2004) and Mura et al. (2009). We use the solar UV photon flux in Ermolli et al (2013) scaled to the orbit of Mercury and we obtain that, at the average distance of 0.38 AU and at the sub-solar point, *A* is approximately equal to $5 \cdot 10^{-5}$ s$^{-1}$. The value of $\Phi_n$ for other regions of the surface of Mercury is obtained by using the well-known cosine law.

The energy distribution of the emitted Na atoms has been extrapolated by laboratory measurements of electron stimulated desorption (200 eV) of adsorbed Na from SiO$_2$ film (Yakshinskiy and Madey, 1999). The data can be modelled with a Weibull function as in Johnson et al. (2002):

$$f(E) = x(1+x) \frac{EU^x}{(E+U)^{2+x}} \qquad (3)$$

where *E* is the energy of the emitted particle, *T* is the temperature of the distribution, *x* is a parameter (here *x*=0.7), and *U* is the characteristic energy of the distribution equal to 0.052 eV.

The problem of the (primordial) abundance of volatile elements of Mercury (i.e. water, alkalis like Na, K, Li, together with the O and S linked to Fe as FeO and FeS, see Goettel, 1988) dates back to the first ground-based observations and to the exploration by the Mariner 10 mission (see Chapman, 1988, and references therein). The pre-MESSENGER data on the surface composition of Mercury as well as the bulk density of the planet could be fitted by a wide range of models, ranging from the ones extremely rich in refractory elements (e.g. Fe, Ni, Mg, Ca, Al, as predicted by the equilibrium condensation theory of the solar nebula, see Goettel, 1988) up to those rich in moderately volatile elements (see above and Goettel, 1988, and references therein), like Mars. On average, the most plausible bulk composition of the planet was considered to be moderately volatile-rich, mostly due to the remixing of material in the inner Solar System during the planetary accretion process (e.g.: Goettel, 1998 and references therein). In fact, recent studies of the formation of the terrestrial planets (e.g.: Righter and O'Brien, 2011 and references therein) showed that a varying fraction of the mass of Mercury could have originated from the region comprised between Mars and the Main Belt. In these studies, the final outcome, in terms of bulk composition of Mercury, depends significantly on the stochastic nature of the accretion process itself and on still poorly-constrained parameters (e.g. the initial orbits of the giant planets and the mass and temperature distribution in the solar nebula). Anyway, at least ~10% of the mass of Mercury could have originated from less volatile-depleted regions of the Solar System (see e.g. Fig. 4 of Righter and O'Brien, 2011).

Presently, data supplied by the MESSENGER spacecraft provided more detailed information on the surface composition of the planet. Specifically, Evans et al. (2012) reported a global, average Na abundance on the surface of Mercury of 2.9±0.1 wt % (i.e. comparable to the 2.3 wt % of terrestrial crustal rocks, Enghag 2007), while the study of the regional distribution of Na by Peplowski et al. (2014) reported a value of 2.6±0.2 wt % at latitudes lower than 60° and a value almost twice as high (4.9±0.7 wt %) for latitudes comprised between 80° and 90°. In addition to the direct measurement of Na abundance, the elevated K/Th and K/U ratios support the idea that Mercury is significantly more volatile-rich than previously thought (Peplowsky et al., 2011; McCubbin et al., 2012). However, the lack of information on the interior state of the planet (e.g.: the presence of a FeS layer in the outermost regions of the core capable of trapping K, U and possibly Th, McCubbin et al., 2012) and on the high-pressure behavior of these tracing species (e.g. their metal/silicate partition coefficient, McCubbin et al., 2012) imposes caution in interpreting these data, as the present surface abundances of K, Th and U could be unrepresentative of the bulk (primordial) ones of Mercury. Finally, the recent MESSENGER's identification of hollows in close proximity to impact craters as possible product of recent loss of volatiles (through some combination of sublimation, space weathering and outgassing, Blewett et al., 2013) supports the idea of a non-negligible presence of volatile elements on the surface of the planet.

The duration of the volcanic activity on the planet is another source of uncertainty about the abundance of volatiles on the surface of Mercury (linked to a limited knowledge of its interior structure and composition). Recent observations by the MESSENGER spacecraft (Prockter et al. 2010) revealed possible evidences that volcanism and impact-induced effusive phenomena extended up to about 1 Ga ago (and potentially even more recently). In fact, thermal models, constrained using the data from the MESSENGER mission (Michel et al. 2013), indicate that the silicatic mantle of Mercury likely was in a molten state over a significant fraction of the life of the planet. The most extreme scenarios indicate, as an upper limit, that convection in the molten mantle could still be acting at present time (Michel et al. 2013). As a lower limit,

Mercury's mantle remained molten and convecting across the first 1 Ga of the life of the planet, i.e., until after the LHB (Michel et al. 2013). If the material composing the mantle of Mercury was not significantly depleted in volatiles with respect to the surface (see previous discussion), even in the latter scenario volcanic phenomena would have brought a new supply of volatile materials to replace those removed in the previous, more active phases of the life of the Sun. This idea received a strong observational confirmation from the results of Peplowski et al. (2014) in studying the regional distribution of Na. The higher abundance of Na at high latitudes discovered is geologically associated to smooth plains of volcanic nature whose crater records indicate that were formed across or after the LHB (Denevi et al. 2013). From the difference in the Na content between younger and older terrain units, it follows that solar-induced depletion of Na contributed to the surface evolution of Mercury until the LHB, and that the abundance of Na was likely significantly higher in more ancient times. Finally, the overall abundance of Na on the surface of Mercury, i.e. the volatile material affected by the PSD, indicates that this process is most likely still active today (Leblanc and Johnson, 2003).

### *3.3 Meteoritic impact vaporization*

Differently from the other processes described in the previous sub-sections, MIV is the only one not directly related to the Sun. The early stages of the Solar System's formation are generally depicted as characterized by high levels of meteoritic bombardment, giving us the vision of a very chaotic and ubiquitous presence of meteoroids of different sizes (from proto-planets to sub-meter sized residuals of the mutual encounters). Only in recent ages we can find more 'regular' populations of meteoroids and with similar orbital elements, even though they are still far from dynamical equilibrium (Borin et al. 2009; 2010).

Hence, the meteoroids present today in the Solar System are mainly the remnants of the big events occurred during the formation of the planets and during the LHB, plus the result of the progressive disruption of the meteoroids themselves due to mutual impacts in space and with planets and satellites. In addition to that, a non-negligible fraction of the dust present in the inner Solar System comes from the mechanical disruption and chemical sublimation of comets along their recurrent orbits as they pass close to the Sun (Nesvorny et al., 2010).

For all these reasons, tracing back in time the evolution in size and velocity distributions of meteoroids in general, and in particular for the ones in the region around Mercury along its history, is a hard task, and models cannot account for the entire and complex history and mixing of objects of different origin, size and velocity that may have occurred at Mercury and at its heliocentric distance as well. Nevertheless, we attempt to provide a reasonable picture of the meteoritic contribution to Mercury's erosion along its history by using the current knowledge in terms of both size and velocity distribution, by supposing that the present situation may depict also past conditions. This approach will give a lower limit estimation of both escape and influx contributions, assumed to be constant from the LHB to present time.

Vaporization induced by meteoritic impact can be derived by using the Cintala (1992) formula:

$$V = \frac{4}{3}\pi \cdot r^3 \cdot \left(c + dv + ev^2\right) \quad (4)$$

where $V$ is the vaporized volume, $r$ the size of meteorites, $v$ its velocity, and $c$, $d$ and $e$ constants depending on both the soil temperature and the projectile material.

A constant precipitation of particles of small size (<100 μm) occurs on the surface of Mercury at a modal velocity of about 20 km/s (Cintala, 1992),; the expected effects are (1) regolith mixing and (2) vaporization of the surface. Higher velocities up to 80 km/s, anyway, can be reached (Marchi et al., 2005; Cremonese et al., 2005; 2006) depending on the size and origin of the meteoroids, and Mercury heliocentric distance. Sporadic, larger objects can cause local density enhancement of all exospheric species, including Na (Mangano et al., 2007), but their contribution to the global Mercurian exosphere is considered to be very uncertain (see for example Killen et al., 2007). In a previous study, Mura et al. (2009), used the same model as this study to estimate the MIV contribution. By considering a thermal velocity distribution at about 2500 K of the ejecta (Eichhorn, 1978) and the precipitating particles uniformly distributed over the surface, they perform simulation which shows that MIV is not the main Na source for the present configuration of Mercury's environment. Conversely, Leblanc and Johnson (2003) come to the opposite conclusion that meteoritic contribution is well comparable with the other processes (up to 20% of the total Na ejecta), leaving the topic an open debate.

## 4. The influence of release processes on the erosion of the Mercury's surface

### 4.1 Ion sputtering

For IS, a Montecarlo model is used to evaluate the number of particles that can escape from the Mercurian exosphere versus epoch phases, thus allowing to estimate the actual surface erosion as a function of age. The model is similar to the one already presented by Mura et al. (2009): it assumes that the magnetic dipole moment in the past was not stronger than now, and then it was unable to shield the more intense solar wind flux. Actually, the magnetic field could have been different in past times; but any different assumption (e.g.: stronger dipole) would be arbitrary, so that we assume as zero-order hypothesis that the dipole moment was constant throughout ages, up to present time.

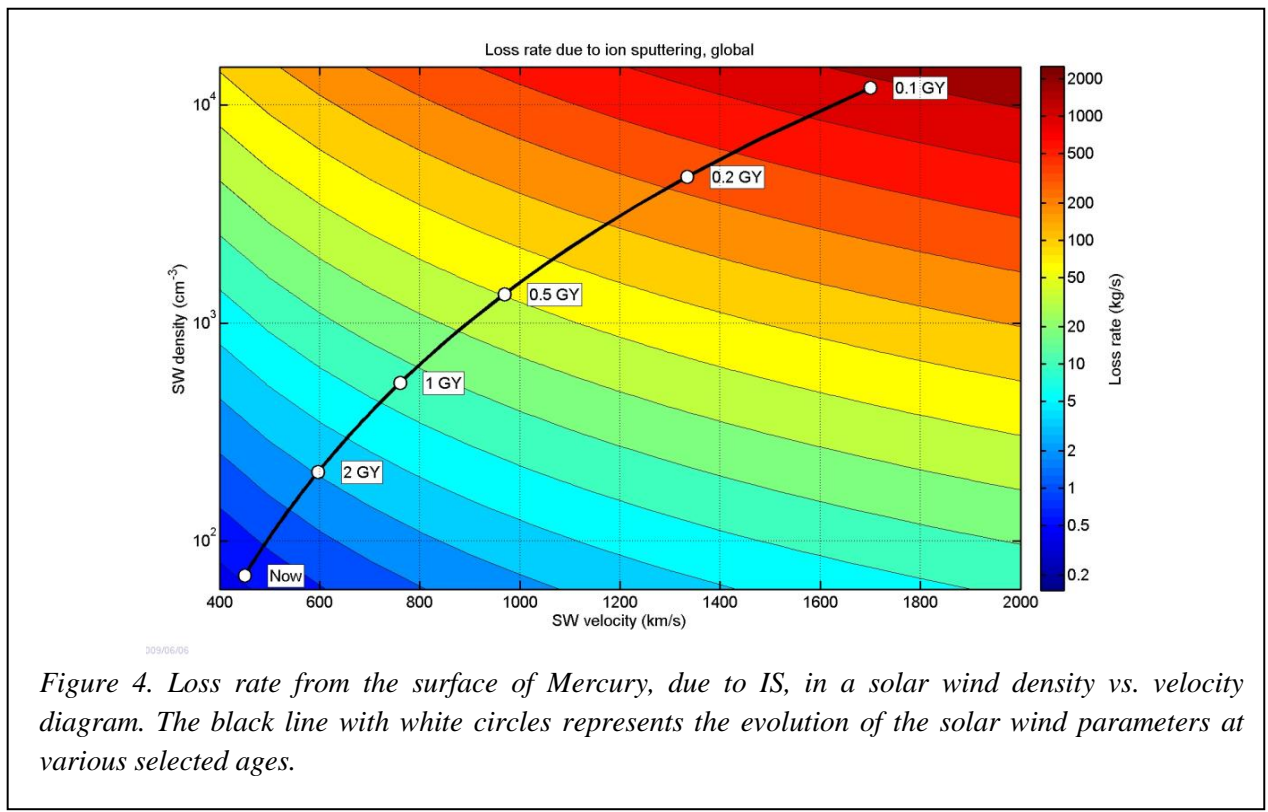

*Figure 4. Loss rate from the surface of Mercury, due to IS, in a solar wind density vs. velocity diagram. The black line with white circles represents the evolution of the solar wind parameters at various selected ages.*

The solar wind velocity and density are variable input parameters, ranging from 400 to 2000 km/s and from 60 to $10^4$ cm$^{-3}$, respectively (see Figure 2 and 3, and Lundin et al., 2007, Figure 5). For each combination of solar wind velocity and density, a model run has been performed with temporal steps of 100 Ma and the total loss rate from the surface has been computed. The surface composition here is not relevant, as we use an average IS yield, which is considered to be constant and uniform, because IS can act on deeper layers of surface. The loss rate from the surface, in kg/s, is plotted in Figure 4, as a function of solar wind velocity and density, whose time variability is derived from Figures 2 and 3. The time variation of the solar wind parameters is also shown. The loss, in early phases, is of the order of ~$10^3$ kg/s.

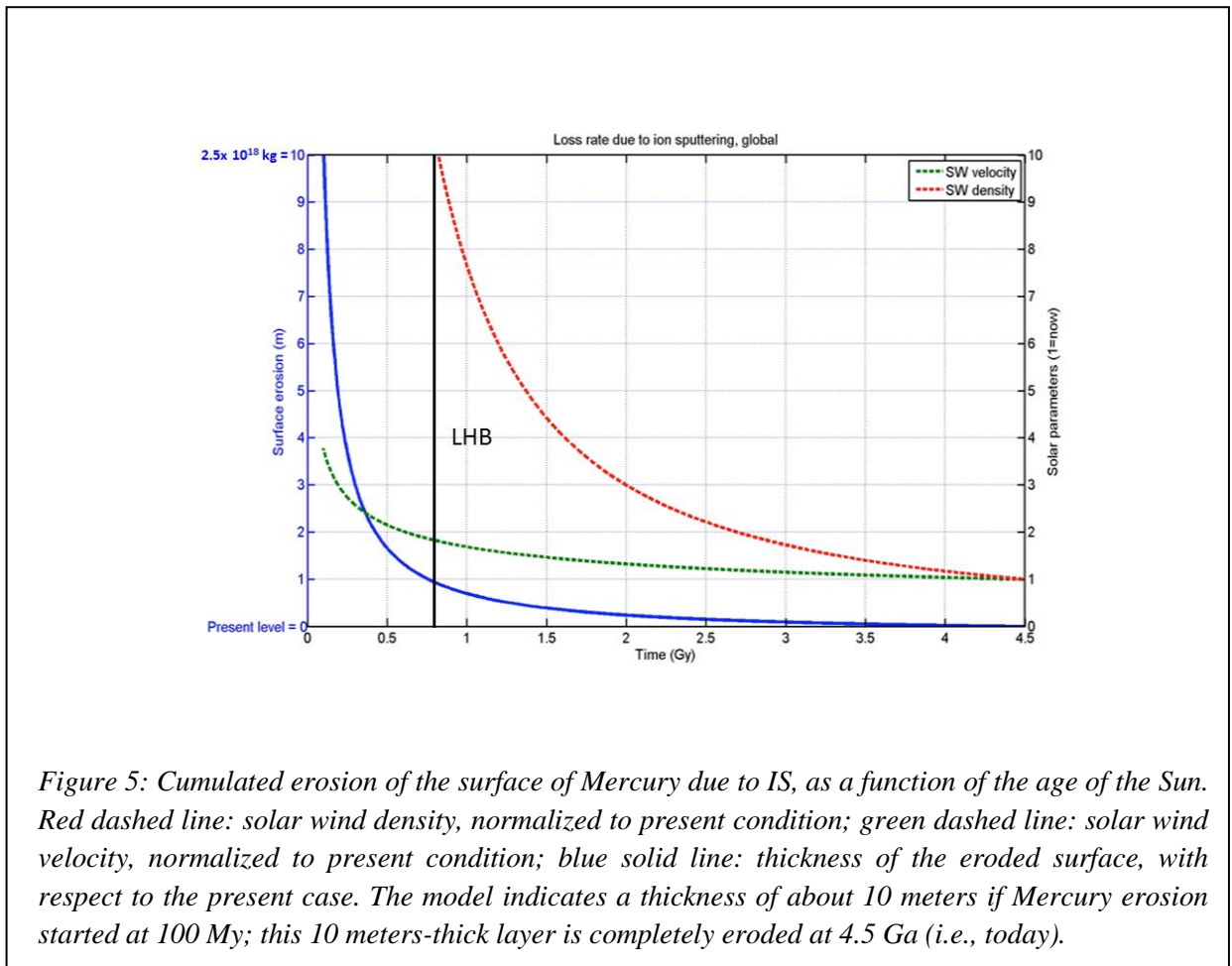

*Figure 5: Cumulated erosion of the surface of Mercury due to IS, as a function of the age of the Sun. Red dashed line: solar wind density, normalized to present condition; green dashed line: solar wind velocity, normalized to present condition; blue solid line: thickness of the eroded surface, with respect to the present case. The model indicates a thickness of about 10 meters if Mercury erosion started at 100 My; this 10 meters-thick layer is completely eroded at 4.5 Ga (i.e., today).*

Data in Figure 4 can be converted from eroded mass to eroded surface thickness by knowing the average regolith density. Then, it is possible to evaluate the total eroded thickness versus time (Figure 5). This is done by integrating the erosion over time along the black curve in Figure 4 (with a numerical integration method). This computation shows that the surface was eroded mainly in early phases up to a value of about 10 meters in total, and that since the LHB Mercury should have lost about 1 m of material from its surface.

*4.2 Meteoritic impact vaporization*

We can consider the actual influx of meteorites onto the surface of Mercury to be divided in two different size-ranges, which are distinctive also of the different origin of the involved projectiles. We define as 'smaller objects' the meteorites smaller than $10^{-2}$ g (equal to $10^{-3}$ m radius, assuming a density of 3.0 g/cm$^3$), which are considered to come mainly from the inner Solar System. They are the result of comets evaporation while passing close to the Sun, and of the 'natural' meteoritic disruption over time due to collisions, as the detection of β-meteoroids from Helios dust experiment suggested for the first time (Grün et al., 1984), and as the ULYSSES dust experiment further confirmed (Mann et al., 2004). 'Bigger objects' are meteoroids in the range $10^{-2}$ - $10^{2}$ m; they mainly originate in the Main Belt (Marchi et al., 2005) and are injected in the inner Solar System due to the effects of the υ6 resonance (Morbidelli and Gladman, 1998).

The total contribution of the process of MIV to the surface of Mercury is given by the net sum of the influx and the outflux of all sizes. In the frame of the current knowledge of the meteoritic population around Mercury, we can try to estimate the total influx and outflux contributions to the planet due to the meteoritic process over time.

Cintala (1992) gave the cumulative production rates of impact melt and vapour generated by 'small objects', stating that the bulk of the contribution comes from the mass range $10^{-8} \div 10^{-2}$ g (that is $10^{-5}$-$10^{-3}$ m). At present time, the average rate of vapour production is calculated by using equation (5) and is equal to $1.41 \cdot 10^{-15}$ g/(cm$^2$ s). This outward flux corresponds to about $1.6 \cdot 10^{7}$ kg/y (or 0.5 kg/s), a value that is of the same order of magnitude as the one ($4.26 \cdot 10^{7}$ kg/y) computed by Cremonese et al. (2006). Also the size range $10^{-2}$-$10^{2}$ m can account for a similar value of produced vapour. Applying the calculation of vaporization given by Cintala (1992), we derive $8.5 \cdot 10^{7}$ kg/y (or 2.7 kg/s). If we assume that the small region in the gap between the two ranges ($10^{-3}$-$10^{-2}$ m radius) does not change significantly these estimations, the total material vaporized due to meteoritic component is of the order of ~kg/s.

A Maxwellian distribution of vaporized particles, as suggested by Eichhorn (1978), would allow only 0.01% of the particles to escape the gravity of Mercury for a typical temperature of 2500 K, and only 1% for 5000 K. Then, the net escape in space is in the range of $10^4 \div 10^6$ kg/y, whereas almost the totality of the vaporized material via MIV falls back on the planetary surface and mostly sticks or remains gravitationally bound to the ground in a dynamical equilibrium.

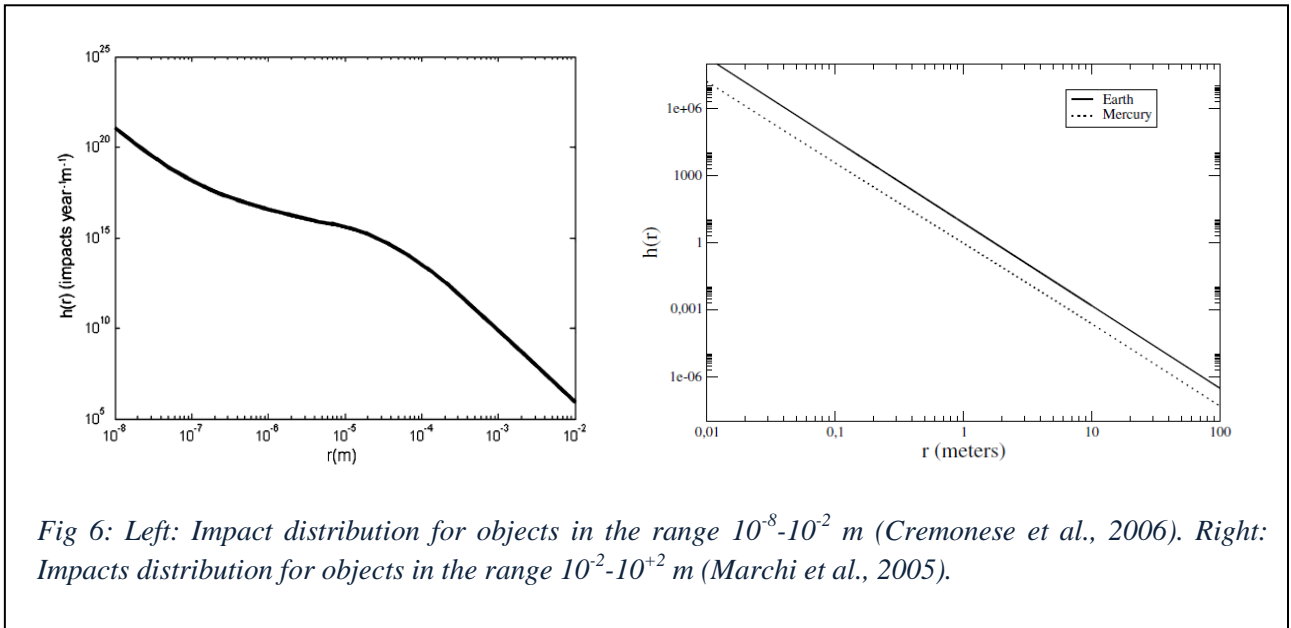

*Fig 6: Left: Impact distribution for objects in the range $10^{-8}$-$10^{-2}$ m (Cremonese et al., 2006). Right: Impacts distribution for objects in the range $10^{-2}$-$10^{+2}$ m (Marchi et al., 2005).*

On the other hand, regarding the calculation of the infalling meteoritic material, Cremonese et al. (2006) plotted the mass distribution *h(m)* of incoming objects (Figure 6, left) as it was calculated by Cintala (1992), and extended the range up to $10^{-2}$ m in radius. The integration of this curve gives the incoming flux due to 'smaller objects' equal to $10^3$ kg/y, or a factor 170 more if we assume that the Borin et al. (2009) calculations apply to the whole size range considered (and not only to $10^{-4}$-$10^{-6}$ m). A similar plot for the range $10^{-2}$ - $10^2$ m (Figure 6, right), as it was derived by Marchi et al. (2005), accounts for an influx of the order of $10^6$ kg/y (or 0.1 kg/s). The comparison between the two values clearly shows that the 'bigger objects' contribution is dominant. This value is in accordance with the 16 tons/day estimated by Mueller et al. (2002), equal to 0.18 kg/s.

In conclusion, the net MIV contribution to the surface of Mercury results in an influx that is bigger or comparable with respect to the loss by evaporation and consequent escape. Hence, the meteoritic impacts are estimated to work in the sense of increasing the mass of Mercury, even if their net accumulation rate to the surface is about 5 cm/Ga. Hence, given its small contribution and the big uncertainties related to the estimations, we can conclude that the MIV contribution over time does not affect significantly the discussion on Mercury surface erosion, whereas a substantial reshaping of the surface composition may be induced by MIV.

*4.3 Photon stimulated desorption*

Here we discuss the evolution of the surface of Mercury in case PSD is acting, although it is limited by the diffusion rate inside the regolith. We underline that the model we apply (described in detail in Mura et al., 2009) refers only to Na. In summary, the model considers that the surface of Mercury has a limited amount of Na, so that the released flux through the PSD process acting upon this species is also limited. From inside, diffusion replenishes the surface concentration; in addition to that, proton precipitation can accelerate the replenishment of fresh Na in the uppermost layers of the surface (ED, Mura et al., 2009; Sarantos et al., 2010). The used Montecarlo model launches some test particles from the surface of Mercury and tracks them until they escape the gravity field or fall back. This model has been already used, and showed very good agreement with Na observations. The loss rate L at present time is 0.1 kg/s and it is proportional to:

$$L = (F_D + F_{ED})\, e \qquad (5)$$

where $F_D$ is the diffusion term, which is proportional to temperature (Killen et al., 2004); $F_{ED}$ is the proton-enhanced diffusion term, proportional to the solar wind flux onto the surface; $e$ is the escape probability. The loss rate in the past is calculated by running several instances of the model with different solar wind parameters and UV flux. The surface temperature map at present is the same as in Killen et al., (2004); the temperature in the past is estimated by assuming that:

$$T^4 \sim F_{UV.} \qquad (6)$$

Because of this complex combination of effects, the actual volatile loss rate L depends on effective resurfacing/remixing processes (like meteoritic impacts or volcanism) able to refill the surface with volatile elements, on the solar wind density and velocity (IS and ED processes), on the solar UV flux (PSD process), on the surface temperature (which controls the diffusion term, see Killen et al., 2004) and finally on the solar radiation pressure, which affects the escape probability. In fact, we know that ~50% of the Na lost in space is due to the radiation pressure effect (Killen et al., 2001, 2004, Mura 2012). Seasonal variations of these parameters (because of the orbital eccentricity) are not considered here, as their effects average over long time periods. Moreover, we don't know much about the evolution of the eccentricity of the orbit over timescales of the order of 1 Ga here considered. Dynamical simulations, including the perturbations of the other planets and the effects of general relativity, indicate that the Mercury eccentricity can vary between 0 and 0.25 over a few Ma and that, superimposed to these faster oscillations, there are secular changes that can rise the average eccentricity of the planet to values higher than 0.3 over $10^8$ years (De Pater & Lissauer, 2010).

The result is shown in Figure 7, and it shows that the surface was depleted in Na mainly in early phases, and that the estimated Na mass loss amounts to about $5 \times 10^{18}$ kg over the last 4.4

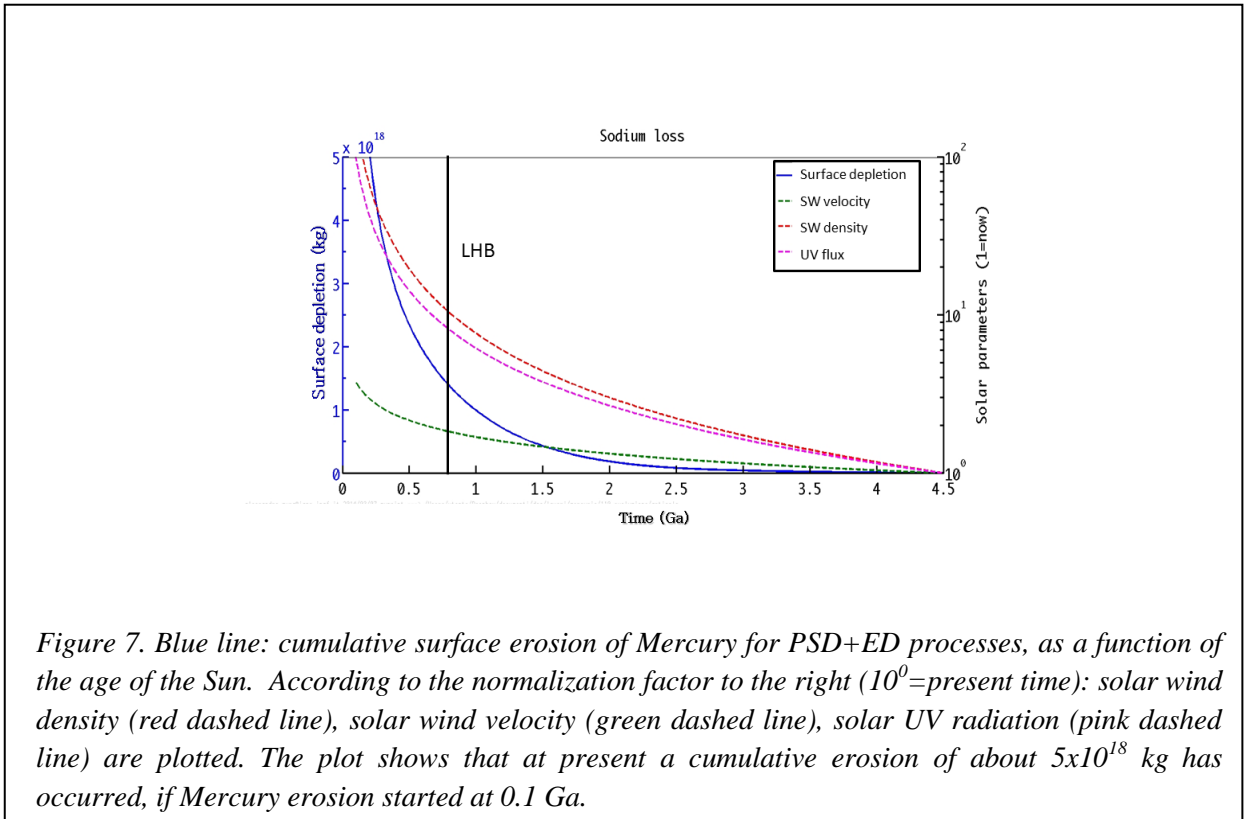

*Figure 7. Blue line: cumulative surface erosion of Mercury for PSD+ED processes, as a function of the age of the Sun. According to the normalization factor to the right ($10^0$=present time): solar wind density (red dashed line), solar wind velocity (green dashed line), solar UV radiation (pink dashed line) are plotted. The plot shows that at present a cumulative erosion of about $5 \times 10^{18}$ kg has occurred, if Mercury erosion started at 0.1 Ga.*

Ga. From the LHB to now, the depletion of Na amounts to about $10^{18}$ kg, i.e. a factor 5 lower.

By considering the previous depleted Na values, we can attempt to figure out the mutual relationship between the possible past Na surface concentration and the related Na-depleted layer depth (see Figure 8). We adopt an approximate surface material density of 2000 kg/m$^3$ (Cintala, 1992), and consider the differential Na concentration (e.g. ± 1 wt%) with respect to the present time measurements (2.4 wt%, i.e. the lowest value measured by Peplowski et al., 2014). In the figure, the blue dashed line refers to the Na removed over the last 4.4 Ga (5x10$^{18}$ kg), the red solid line refers instead to the last 3.7 Ga (1x10$^{18}$ kg), i.e.: after the LHB; the depleted concentration ranges from 0 wt%, which is the asymptotic limit for the related depth, up to 4.6 wt%, i.e. the differential concentration of Na needed to bring the initial Na content to the cosmic concentration of 7%, once we assume the present 2.4 wt % as an offset. The values reported in Figure 8 assume the unrealistic end-member case that implies the existence of a continuous diffusion of Na to the surface layer affected by PSD+ED from the underlying bedrock. The effective range of Na-depleted depths strongly depends on how the remixing processes are effective and on the thickness of the layer they affect. It is likely that, in the size range considered, meteoritic impacts (i.e. "Population 2" impactors, Strom et al. 2005) were able to remix a few hundred meters on a global scale starting from LHB, bringing Na-undepleted material to the surface. Before the LHB both volcanic activity and "Population 1" meteoritic impacts could have replenished the Na content within a layer of the order of kilometers. For what it concerns volcanic activity, we recall that the depth of magma chambers on Earth is estimated to be of the order of 10 km (e.g.: Becerril et al., 2013 and references therein) and that the thickness of a single lava flow on the Earth and the Moon can range from a few meters to a few tens of meters (Hulme, 1974). For what it concerns the role of impacts, instead, we recall that "Population 1" impactors are extracted from the asteroid belt by size-independent processes, therefore producing larger impactors than those of "Population 2", and capable of excavating material from depths of a few km (e.g. a projectile of 1 km in diameter impacting Mercury at 20 km/s would excavate its surface to depths of 1.5-2 km), and that "Population 1" plausibly affected the surface of Mercury at a global level (Strom et al., 2005). As a consequence, the lowest Na-depleted concentration value is determined by such effectiveness (about 1 wt% including LHB age, as shown in Figure 8), whereas the highest values go up to 4.6 wt %. It is noticeable that in both the two considered profiles, the solar radiation seems to have been able to cause Na concentration loss from primordial state, down to present times value, by means of deep remixing, ED, and finally PSD.

The fact that the largest fraction (about 80%) of the Na depletion occurs during the first

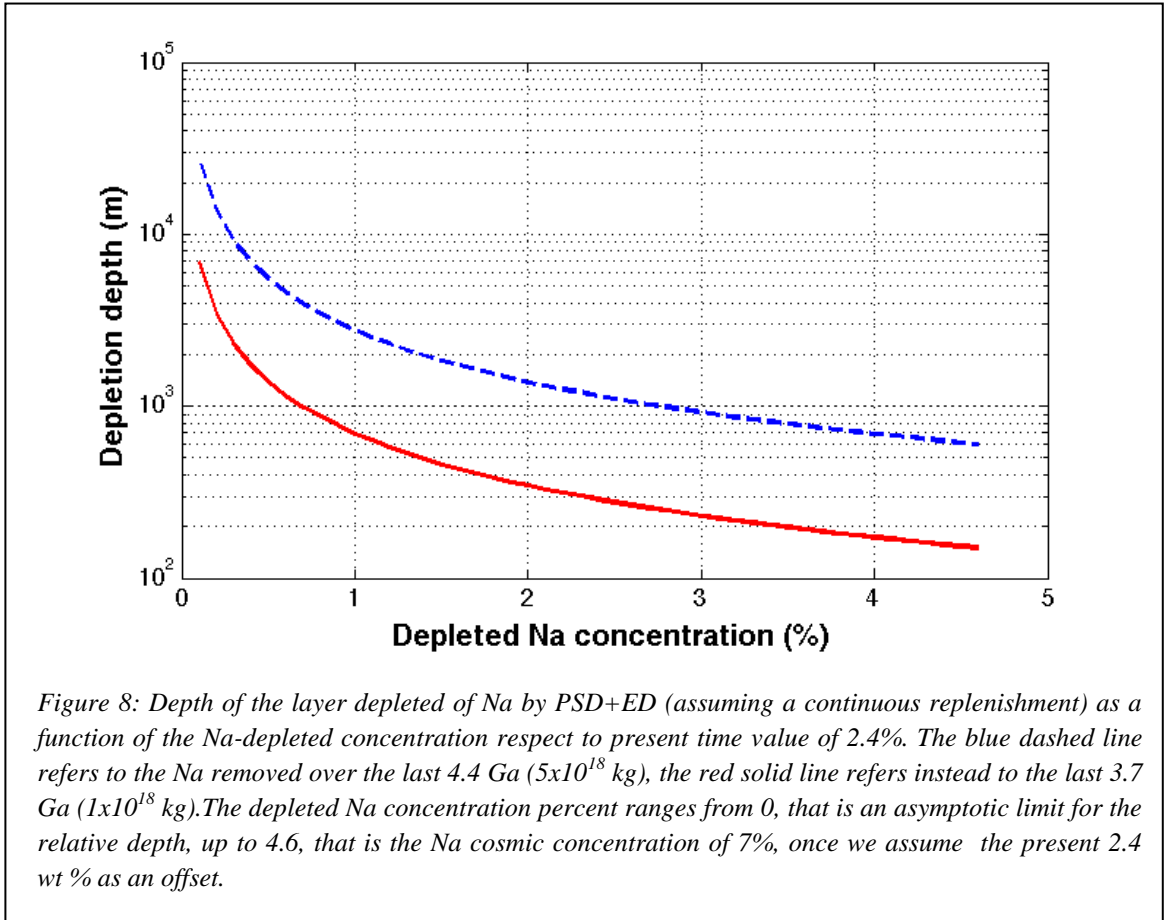

*Figure 8: Depth of the layer depleted of Na by PSD+ED (assuming a continuous replenishment) as a function of the Na-depleted concentration respect to present time value of 2.4%. The blue dashed line refers to the Na removed over the last 4.4 Ga ($5x10^{18}$ kg), the red solid line refers instead to the last 3.7 Ga ($1x10^{18}$ kg). The depleted Na concentration percent ranges from 0, that is an asymptotic limit for the relative depth, up to 4.6, that is the Na cosmic concentration of 7%, once we assume the present 2.4 wt % as an offset.*

0.5-0.7 Ga of the life of the Solar System (see Figures 7 and 8) is consistent with the observational evidence that on Mercury terrains more ancient than the LHB are about a factor two less rich in Na than the post-LHB terrains (Peplowski et al., 2014).

## 5. Discussion and conclusions

In Table 1 we give a summary of our results in terms of mass loss and erosion (where applicable).

| Process | From 4.4 Ga ago | | From 3.7 Ga ago | | Present | |
|---|---|---|---|---|---|---|
| | Loss, kg | Erosion, m | Loss, kg | Erosion, m | Loss rate, kg/s | Erosion rate, cm/Ga |
| IS | $5 \times 10^{18}$ | 20 | $4 \times 10^{17}$ | 2 | 0.2 | 5 |
| MIV | $-4 \times 10^{15} \div 0$ | $-3 \times 10^{-2} \div 0$ | $-3 \times 10^{15} \div 0$ | $-2 \times 10^{-2} \div 0$ | $-0.3 \div 0$ | $-5 \div 0$ |
| PSD - ED | $5 \times 10^{18}$ | N.A. | $1 \times 10^{18}$ | N.A. | 0.1 | N.A. |

*Table 1: Mass loss and erosion rates (N.A. = not applicable), according to different time intervals. IS=Ion Sputtering; MIV=Meteorite Impact Vaporization; PSD=Photon Stimulated Desorption; ED= Enhanced Diffusion.*

As derived in the previous sections, the surface erosion of Mercury, due mainly to IS, from the T-Tauri phase of the Sun to now, could amount to a shell about 10 meters thick. At the same time, the combined processes of PSD and ED would have removed from Mercury sufficient Na (and, analogously, volatile species like K) to deplete a crustal shell up to a depth of a few kilometers. In fact, the solar radiation seems to have been able to cause Na concentration loss from primordial state (before as well as after LHB), down to present times measurements, by means of deep remixing processes, ED, and PSD combined effects.

About 90% of the surface erosion and 80% of the Na depletion would have occurred across the first 0.7-0.8 Ga from the condensations of Ca-Al-rich inclusions found in chondritic meteorites, therefore before and during the LHB. Across this timespan, the Moon and the terrestrial planets were still characterized by an active geophysical evolution, therefore the resurfacing caused by volcanic activity and/or impact-triggered effusive phenomena (e.g. the lunar maria) would have erased or masked the signatures of this Sun-induced erosive process of the surface and replenished the latter of Na (and K), as testified by the regional distribution of Na on Mercury (Peplowski et al., 2014) once considered in its geologic context (Denevi et al., 2013). Moreover, it has been suggested that the LHB erased the previous cratering records on the surfaces of the Moon and the terrestrial planets (Strom et al., 2005), again acting against the possibility to unequivocally identify signatures of the primordial erosion of the surface of Mercury caused by the solar wind and the solar radiation. However, the Sun-induced erosive process continued also after the LHB, stripping to the surface of Mercury about 1 m of material during the last 3.7 Ga, and removing an amount of Na equivalent to depleting a shell a few hundreds meters thick. At the same time, the impact rate in the inner Solar System dropped down significantly with the passage from 'Population 1' to 'Population 2' of impactors. Hence, such a Sun-induced erosion of the surface of Mercury needs to be taken into account to properly interpret the crater record and the geological history of the planet. In fact, even if the erosion caused by solar wind and radiation from the LHB to now appears quite limited, nevertheless it could have contributed to shape the morphology of the Mercury landscape in several ways. The secular erosion of the crater rims, especially where their slope approaches the angle of repose of the regolith, could have triggered landslides and mass movements that could cause these craters to appear older and more degraded than their real age. This effect would be more marked for small craters (i.e. with diameters of the order of a few tens of meters). All these effects would plausibly be more marked in geologic and impact structures dating before the LHB than in younger structures, but it is likely that other processes (e.g. those responsible for crater degradation) played a larger role in affecting older structures. Finally, Sun-induced erosion could also cause the shrinking of the areas affected by the ejecta blankets or the deposition of the material (as solid fragments or melt droplets) of the impactors, since those regions where the thickness of the layer of deposited material is inferior or comparable to the secularly eroded thickness would effectively disappear, thus leading to a misinterpretation of the impact event that generated them. Such an effect could be in principle responsible for the absence on the surface of Mercury of swirls (Blewett et al., 2010) like those observed on the Moon (e.g.: Neish et al., 2011 and references therein). Lunar swirls are sinuous features characterized by optically bright albedo that have been observed both in the lunar maria and highlands and that are associated to regions of enhanced magnetic field intensity in the lunar crust (see previous references). As the spectral properties of the material composing the swirls are similar to those of immature regolith (Blewett et al., 2013), it has been suggested that the anomalies in the magnetic field protect the swirls from the effects of space weathering (Hood & Schubert 1980). The observations of the Mini-RF synthetic aperture radar on-board the Lunar Reconnaissance Orbiter revealed that swirls are a very surficial phenomenon, whose thickness does not exceed a few decimeters (Neish et al. 2011). The higher surface erosion rate at Mercury due to the more intense solar wind environment would result in quite limited (i.e. of the order of a few $10^8$ years) survival times of swirl-like features. These possibilities are intriguing but presently only speculative, expecially for what it concerns the swirls as their origin is still not fully understood (see e.g. Blewett et al., 2010; Neish et al., 2011, and references therein): more detailed investigations are needed in order to assess the interplay between the resurfacing and the erosion due to impact events and those due to solar wind. This study, however, shows that the Mercury environment could be uniquely shaped by processes not observed on the other bodies of the Solar System, since they are more distant from the Sun, and furthermore many terrestrial bodies are shielded by denser atmospheres and/or by strong internal magnetic fields. This peculiar reshaping process needs to be taken into account for properly interpreting the data that the MESSENGER mission is supplying and to pave the road for the investigations that the BepiColombo mission will perform.

## 6. Final remarks

The present study is an attempt to trace back the present surface erosion and volatile depletion rate to the very early stages of the life of Mercury, back to 4.4 Ga ago. The LHB, which is assumed to have occurred between 3.9 and 3.7 Ga ago, falls inside the time span we investigate in this work. It is suggested that the LHB would have erased most pre-existing geological features: hence, we also focused on the total surface erosion and Na depletion that occurred during the last 3.7 Ga only.

The erosion due to IS amounts to 10 m across the last 4.4 Ga, and 1 m since the end of LHB. These values are limited, but they can however have implications for the surface evolution of Mercury and be possibly related to the lack of features analogous to the lunar swirls (should they actually have formed on Mercury). There is no erosion due to MIV, since the net effect seems to be a minor mass increase due to the precipitation of meteoroids.

The estimation of the depletion rate due to the combined effects of PSD and ED is computed by means of a model able to reproduce the observed Na exospheric density with good accuracy. We find that, over the whole timespan considered in this study, these processes could have been able to remove enough Na and K to deplete the crust of Mercury down to depths of a few kilometres from solar T-Tauri phase, and hundreds of meters after the LHB. Nevertheless, these evaluations should be interpreted as an actual potential effect, subject to several boundary constraints and uncertainties which could affect these values and limited by the real efficiency of the processes replenishing the surface of Mercury of these elements.

In the limit of the mentioned uncertainties, we may conclude that the overall loss rate from the surface of Mercury via all Sun-related surface erosion and volatile depletion processes are important elements for the interpretation of the surface features. It is also worth to note that, since the Sun-induced surface erosion and volatile depletion was not constant over time, the identification of such effects on different surface features could help in estimating the age of their formation.

**Aknowledgements**

This paper is financially supported by the Italian Space Agency (ASI) under contract 'SERENA', number I/090/06/0.